\documentclass[12pt]{article}
\usepackage{amssymb,amsmath}
\usepackage[dvips]{graphicx}

 1
\makeatletter
\newcommand{\bea}{\begin{eqnarray}}
\newcommand{\eea}{\end{eqnarray}}
\makeatother
\begin{document}
%
\begin{center}
{\Large\bf 
 Hidden particle production at the ILC \\
}
\end{center}
\vspace{8mm}
\begin{center}
\normalsize
{\large \bf 
Keisuke Fujii$^{(a)}$
 \footnote{E-mail: keisuke.fujii@kek.jp}, 
Hitoshi Hano$^{(b)}$
 \footnote{E-mail: hano@icepp.s.u-tokyo.ac.jp},  
Hideo Itoh$^{(a)}$
 \footnote{E-mail: hideo@post.kek.jp},  
Nobuchika Okada$^{(a)}$ 
 \footnote{E-mail: okadan@post.kek.jp},  
 and 
Tamaki Yoshioka$^{(b)}$
 \footnote{E-mail: tyosioka@icepp.s.u-tokyo.ac.jp}
}
\end{center}
\vskip 1.2em
\begin{center}
${}^{(a)}$ {\it 
High Energy Accelerator Research Organization (KEK) \\ 
1-1 Oho, Tsukuba 305-0801, Japan } 

${}^{(b)}$ {\it 
University of Tokyo, 
International Center for Elementary Particle Physics (ICEPP) \\ 
7-3-1 Hongo, Bunkyo-ku, Tokyo 113-0033, Japan}
%
\end{center}
\vskip 1.0cm
\begin{center}
{\large Abstract}
\vskip 0.7cm
\begin{minipage}[t]{16cm}
\baselineskip=19pt
\hskip4mm
In a class of new physics models, 
 new physics sector is completely or partly hidden, 
 namely, singlet under the Standard Model (SM) gauge group. 
Hidden fields included in such new physics models 
 communicate with the Standard Model sector 
 through higher dimensional operators. 
If a cutoff lies in the TeV range, 
 such hidden fields can be produced at future colliders. 
We consider a scalar field as an example of the hidden fields. 
Collider phenomenology on this hidden scalar is similar to 
 that of the SM Higgs boson, but there are several 
 features quite different from those of the Higgs boson. 
We investigate productions of the hidden scalar 
 at the International Linear Collider (ILC) 
 and study the feasibility of its measurements, 
 in particular, how well the ILC 
 distinguishes the scalar from the Higgs boson, 
 through realistic Monte Carlo simulations. 
\end{minipage}
\end{center}
\newpage
%

\section{Introduction}
In a class of new physics models, 
 a new physics sector is completely or partly singlet 
 under the Standard Model (SM) gauge group, 
 SU(3)$_C \times$SU(2)$_L\times$U(1)$_Y$. 
Such a new physics sector, 
 which we call ``hidden sector'' throughout this paper, 
 includes some singlet fields. 
These hidden sector fields, in general, couple with 
 the SM fields through higher dimensional operators. 
If the cutoff scale of the higher dimensional operators 
 lies around the TeV scale, 
 effects of the hidden fields are accessible 
 at future colliders 
 such as the Large Hadron Collider (LHC) 
 and the International Linear Collider (ILC).

There have been several new physics models proposed 
 that include hidden fields. 
The most familiar example would be the Kaluza-Klein (KK) modes 
 of graviton in extra dimension scenarios \cite{ADD} \cite{RS}. 
A singlet chiral superfield 
 in the next to Minimal Supersymmetric Standard Model (MSSM) 
 \cite{NMSSM} is also a well-known example, 
 which has interesting implications, in particular, 
 on Higgs phenomenology in collider physics \cite{Dobrescu:2000jt}.  
Another example is the supersymmetry breaking sector 
 of the model proposed in Ref.~\cite{IOY}, 
 where a singlet scalar field couples with the SM fields 
 through higher dimensional operators 
 with a cutoff around $\Lambda=1-10$ TeV and 
 its collider phenomenology at the LHC and ILC 
 has been discussed. 
A very recently proposed scenario \cite{unparticle}, 
 ``unparticle physics'', belongs to this class of models, 
 whose phenomenological aspects have been intensively 
 studied by many authors. 
%

In this paper, we introduce a hidden scalar field 
 and investigate the hidden scalar production at the ILC. 
We assume that the hidden scalar couples with 
 only the SM gauge fields through higher dimensional operators 
 suppressed by a TeV-scale cutoff. 
In this case, at the ILC, this hidden scalar can be produced 
 through the similar process to the SM Higgs boson production 
 and with the production cross sections comparable to the Higgs boson one. 
Thus, the hidden scalar production has interesting implications 
 on the Higgs phenomenology. 
The crucial difference of the hidden scalar from the Higgs boson 
 lies in that the hidden scalar has nothing to do with 
 the electroweak symmetry breaking. 
This feature reflects the fact that the hidden scalar couples 
 with mostly the transverse mode of the weak gauge bosons while the Higgs boson couples 
 with mostly their longitudinal modes. 
The hidden scalar could be discovered at the LHC 
 (together with the Higgs boson) and then, would be 
 identified as a Higgs boson-like particle. 
It is an interesting issue how to distinguish 
 the scalar irrelevant to the electroweak symmetry breaking 
 from the relevant one (the Higgs boson). 
It would be challenging to tackle this issue with the LHC. 
In this paper, 
 based on realistic Monte Carlo simulations, 
 we study the feasibility of measurements 
 for the hidden scalar productions and its couplings 
 to the SM particles, and show how well 
 the hidden scalar can be distinguished 
 from the Higgs boson at the ILC.

This paper is organized as follows. 
In the next section, 
 we introduce our theoretical framework 
 and present formulas relevant to our studies. 
In Sec.~3, we show the results from our Monte Carlo simulations 
 and demonstrate how accurately the ILC can measure 
 the typical features of the scalar and distinguish it 
 from the Higgs boson. 
The last section is devoted to summary and discussions.

\section{Theoretical framework}
In this paper, we introduce a real scalar field $\chi$ 
 as a hidden field, which communicates with the SM sector 
 through interactions of the form, 
\bea 
 {\cal L}_{\rm int} = 
 \frac{c_i}{\Lambda^{d_{\rm SM}-3}} \chi \; {\cal O}_{\rm SM}^i , 
 \label{interaction}
\eea
where $c_i$ is a dimensionless coefficient, 
 $\Lambda$ is a cutoff scale, 
 and ${\cal O}_{\rm SM}^i$ is an operator of the SM fields 
 with mass dimension $d_{\rm SM}$. 
We consider the case that the cutoff, 
 which is naturally characterized by a new physics scale, 
 is around the TeV scale. 
As an example, it would be easy to imagine a model like 
 the large extra-dimension models \cite{ADD} 
 whose fundamental scale is in the TeV range 
 or a model with warped extra dimensions \cite{RS} 
 where the effective cutoff scale is warped down to 
 the TeV range from the 4-dimensional Planck scale. 
For a more concrete example, see Ref.~\cite{IOY}. 
In these models, the above effective interaction 
 can be introduced at tree level.

The theoretical requirements for the SM operator 
 ${\cal O}_{\rm SM}^i$ are that it should be 
 a Lorentz scalar operator and be singlet under the SM gauge group. 
Although there are many possibilities for such operators, 
 we assume that the hidden scalar couples with only 
 the SM gauge bosons through the operators descried as follows\footnote{
In fact, it is easy to construct a simple model 
 which can realize this situation. 
We give comments on this respect in the last section. }: 
\bea
 {\cal L}_{\rm int} = 
  -\frac{1}{2} \sum_A\; c_A\; \frac{\chi}{\Lambda} \;
  {\rm tr}\left[
   {\cal F}_A^{\mu \nu}{\cal F}_{A \mu \nu} \right],   
\label{intgauge1} 
\eea 
where $c_A$ is a dimensionless parameter, 
 and ${\cal F}_A$'s ($A=1,2,3$) are the field strengths  
 of the corresponding SM gauge groups, 
 U(1)$_Y$, SU(2)$_L$, and SU(3)$_C$. 
After the electroweak symmetry breaking, 
 Eq.~(\ref{intgauge1}) is rewritten as interactions 
 between $\chi$ and gluons, photons, $Z$- and $W$-bosons. 
\bea 
{\cal L}_{\rm int} = 
 &-& \frac{c_{gg}}{4} 
 \frac{\chi}{\Lambda} G^{a \mu \nu} G^a_{\mu \nu} 
  - \frac{c_{WW}}{2} \frac{\chi}{\Lambda} 
      W^{+ \mu \nu} W^{-}_{\mu \nu}     
   - \frac{c_{ZZ}}{4} \frac{\chi}{\Lambda} 
      Z^{\mu \nu} Z_{\mu \nu}     \nonumber \\  
&-& \frac{c_{\gamma \gamma}}{2} \frac{\chi}{\Lambda} 
   F^{\mu \nu} F_{\mu \nu}     
 - \frac{c_{Z \gamma}}{4} \frac{\chi}{\Lambda} 
   Z^{\mu \nu} F_{\mu \nu}   , 
\label{intgauge2}
\eea
where $G^{a \mu \nu}$, $W^{+ \mu \nu}$, $Z^{\mu \nu}$ 
 and $F^{\mu \nu}$ are the field strengths 
 of gluon, $W$-boson, $Z$-boson and photon, respectively. 
The couplings $c_{gg}$ etc. can be described 
 in terms of the original three couplings, 
 $c_1$, $c_2$ and $c_3$, and the weak mixing angle $\theta_w$ 
 as follows: 
\bea 
 c_{gg} &=& c_3, \nonumber \\
 c_{WW} &=& c_2, \nonumber \\
 c_{ZZ}  &=& c_1 \sin^2\theta_w + c_2 \cos^2\theta_w , 
 \nonumber \\
 c_{\gamma \gamma} &=& c_1 \cos^2\theta_w + c_2 \sin^2\theta_w , 
 \nonumber \\
 c_{Z \gamma} &=& (-c_1+c_2) \sin \theta_w  \cos \theta_w .
\eea 

The hidden scalar can be produced at the ILC 
 through these interactions. 
The dominant $\chi$ production process 
 is the associated production, 
 $e^{+} e^{-} \rightarrow \gamma^*,Z^* \rightarrow Z \chi$  
 and  $e^{+} e^{-} \rightarrow \gamma^*,Z^* \rightarrow \gamma \chi$. 
First, let us consider the process $e^+ e^- \rightarrow Z \chi$. 
The cross section is calculated as  
\begin{eqnarray} 
& & \frac{d \sigma}{d \cos \theta} (e^+e^- \rightarrow Z \chi) 
 =  \frac{1}{68 \pi s}  \sqrt{\frac{E_Z^2-m_Z^2}{s}} 
\nonumber \\
& \times & 
\left[ 
\left( 
 c_{ZZ}^2 (g_L^2+g_R^2) \left( \frac{s}{s-m_Z^2} \right)^2 
 - c_{ZZ} c_{Z \gamma} (g_L+g_R) e 
  \left( \frac{s}{s-m_Z^2} \right) 
  +c_{Z \gamma}^2 e^2  \right) 
 \frac{E_Z^2}{\Lambda^2} (1+\cos^2 \theta) 
\right. 
 \nonumber \\
&+&
\left.
\left( 
 c_{ZZ}^2 (g_L^2+g_R^2) \left( \frac{s}{s-m_Z^2} \right)^2 
 - \sqrt{2} c_{ZZ} c_{Z \gamma} (g_L+g_R) e 
  \left( \frac{s}{s-m_Z^2} \right) 
  + \frac{c_{Z \gamma}^2 e^2}{2}  \right) 
 \frac{m_Z^2}{\Lambda^2} \sin^2 \theta
\right], 
\label{dcrossZ}
\end{eqnarray} 
where $\cos \theta$ is the scattering angle of 
 the final state $Z$-boson, 
 $g_L= 2 (m_Z/v) (-1/2+ \sin^2 \theta_w)$, 
 $g_R= 2 (m_Z/v) \sin^2 \theta_w$, and 
 $E_Z= \frac{\sqrt{s}}{2} \left( 1+ \frac{m_Z^2 - m_\chi^2}{s} \right) $. 
It is interesting to compare this $\chi$ production process 
 to the similar process of the associated Higgs production 
 (Higgsstrahlung), $e^+ e^- \rightarrow Z h$, 
 through the Standard Model interaction 
 ${\cal L}_{\rm int}= \frac{m_Z^2}{v} h Z^\mu Z_\mu$. 
In Figure~1, we show the ratio of the total cross sections 
 between $\chi$ and Higgs boson productions 
 as a function of $\Lambda$ at the ILC with 
 the collider energy $\sqrt{s} = 500$ GeV. 
Here we have taken $c_1=c_2$ and $m_\chi=m_h=120$ GeV. 
The ratio, 
 $\sigma(e^+e^-\rightarrow Z \chi)/\sigma(e^+e^-\rightarrow Z h)$, 
 becomes one for $\Lambda_{\rm IR} \simeq 872$ GeV, 
 and it decreases proportionally to $1/\Lambda^2$. 
Note that in the high energy limit, the $\chi$ production cross
 section becomes energy-independent. 

The coupling manner among $\chi$ and the $Z$-boson pair 
 is different from that of the Higgs boson. 
As can be understood from Eq.~(\ref{intgauge2}), 
 $\chi$ couples with the transverse modes of the $Z$-bosons, 
 while the Higgs boson mainly couples with the longitudinal modes. 
This fact reflects into the difference of the angular distribution 
 of the final state $Z$-boson. 
In the high energy limit, 
 we find $\frac{d \sigma}{d \cos \theta} (e^+e^- \rightarrow Z \chi) 
   \propto 1+ \cos^2 \theta$, 
 while $\frac{d \sigma}{d \cos \theta} (e^+e^- \rightarrow Z h) 
   \propto 1- \cos^2 \theta$. 
Figure~2 shows the angular distributions of the associated 
 $\chi$ and Higgs boson productions, respectively. 
Even if $m_\chi = m_h$ and the cross sections 
 of $\chi$ and Higgs boson productions are comparable, 
 the angular dependence of the cross section 
 can distinguish the $\chi$ production from the Higgs boson one.

The formula for the process 
 $e^{+} e^{-} \rightarrow \gamma^*, Z^* \rightarrow \gamma \chi$ 
 can be easily obtained from Eq.~(\ref{dcrossZ}) 
 for $Z$-boson by the replacements: 
 $c_{ZZ} \to c_{Z \gamma}$, 
 $c_{Z \gamma} \to c_{\gamma \gamma}$ 
 and $m_Z \rightarrow 0$. 
As a result, the cross section of the process 
 $e^+ e^- \rightarrow \gamma \chi$ is found to be
\begin{eqnarray} 
& & \frac{d \sigma}{d \cos \theta} (e^+e^- \rightarrow \gamma \chi) 
 =  \frac{1}{128 \pi s}  \sqrt{\frac{E_\gamma^2}{s}} 
\times 
\nonumber \\
&&\left( 
 c_{Z \gamma}^2 (g_L^2+g_R^2) \left( \frac{s}{s-m_Z^2} \right)^2 
 - c_{Z \gamma} c_{\gamma \gamma} (g_L+g_R) e 
  \left( \frac{s}{s-m_Z^2} \right) 
  +c_{\gamma \gamma}^2 e^2  \right) 
 \frac{E_\gamma^2}{\Lambda_{\rm IR}^2} (1+\cos^2 \theta),  
\label{dcrossGamma}
\end{eqnarray} 
where $E_\gamma = \frac{\sqrt{s}}{2} \left( 1- \frac{ m_\chi^2}{s} \right) $. 
For example, 
 $\sigma (e^+e^- \rightarrow \gamma \chi) = 105~{\rm fb}$ 
 at $\sqrt{s}=500$ GeV with the parameter set: 
 $m_\chi=120$ GeV, $c_1=c_2=1$ and $\Lambda=1$ TeV. 
For the Higgs production, 
 the process $e^{+} e^{-} \rightarrow \gamma^* \rightarrow \gamma h$ 
 is negligible.

Next, we consider $\chi$ decay processes 
 into a pair of gauge bosons. 
Partial decay widths are given by 
\begin{eqnarray} 
\Gamma (\chi \rightarrow g g )
&=& \frac{c_{gg}^2}{8 \pi} 
    \frac{m_\chi^3}{\Lambda^2},  
 \nonumber \\ 
\Gamma (\chi \rightarrow \gamma \gamma)
&=& \frac{ c_{\gamma \gamma}^2}{64 \pi} 
    \frac{m_\chi^3}{\Lambda^2} , 
 \nonumber \\ 
\Gamma (\chi \rightarrow Z Z)
&=& \frac{c_{ZZ}^2}{512 \pi} 
 \frac{m_\chi^3}{\Lambda^2} 
 \; \beta_Z \left(
 3 + 2 \beta_Z^2 +3 \beta_Z^4\right) ,
 \nonumber \\
\Gamma (\chi \rightarrow W W)
 &=& \frac{c_{WW}^2}{256 \pi} 
 \frac{m_\chi^3}{\Lambda^2} 
 \; \beta_W \left( 
 3 + 2 \beta_W^2 +3 \beta_W^4 \right) ,  \nonumber \\
\Gamma (\chi \rightarrow \gamma Z)
&=& \frac{c_{Z \gamma}^2}{128 \pi} \tan^2(2 \theta_w) 
 \frac{m_\chi^3}{\Lambda^2} 
 \; \left( 1-\frac{m_Z^2}{m_\chi^2} \right)^3 ,
\label{width-gauge} 
\end{eqnarray} 
where $\beta_Z = \sqrt{1-4 (m_Z/m_\chi)^2} $, 
 and $\beta_W = \sqrt{1-4 (m_W/m_\chi)^2}$. 
In Figure~3. the branching ratio of the $\chi$ decay is depicted. 
We see that the branching ratio of the $\chi$ decay 
 is quite different from that of the Higgs boson. 
In particular, the branching ratio of 
 $\chi \rightarrow \gamma \gamma$ can be large, 
 ${\rm Br}(\chi \rightarrow \gamma \gamma) \simeq 0.1$ 
 for the parameter set in Figure~3. 
On the other hand, 
 the branching ratio of the Higgs boson 
 into two photons in the SM is at most $10^{-3}$, 
 since the coupling between the Higgs boson and 
 two photons are induced through one-loop 
 radiative corrections. 

There are several models where the branching ratio 
 of the Higgs boson into two photons is enhanced 
 due to new physics effects. 
For example, in the MSSM with a large $\tan \beta$ \cite{MSSMLtan}, 
 the lightest Higgs boson almost coincides with 
 the up-type Higgs boson of the weak eigenstate. 
As a result, the Yukawa coupling to bottom quark is suppressed 
 and two-photon branching ratio is relatively enhanced. 
Another example is the Next to MSSM (NMSSM), 
 where a pseudo scalar ($A^0$) couples to 
 the lightest (SM-like) Higgs boson. 
In this model, the Higgs boson can decay into 
 two pseudo scalars ($h \rightarrow A^0 A^0$) 
 with a sizable branching ratio. 
If the pseudo scalar is extremely light 
 (lighter than twice the pion mass), 
 it dominantly decays into two photons 
 ($A^0 \rightarrow \gamma \gamma$), 
 so that Higgs boson decays into four photons. 
Since the pseudo-scalar is very light, 
 two photons produced in its decay are almost collinear 
 and will be detected as a single photon \cite{Dobrescu:2000jt}. 
As a result, the Higgs decay into two pseudo-scalars, 
 followed by $A^0 \rightarrow \gamma \gamma$, 
 effectively enhances the Higgs branching ratio 
 into two photons \cite{Dobrescu:2000jt}. 
Therefore, the anomalous branching ratio alone is 
 not enough to distinguish such a Higgs boson from $\chi$ 
 (in the associated production with a $Z$-boson) 
 and the measurements of angular distribution 
 and polarization of the final state $Z$-boson are crucial.

Here, let us consider current experimental constraints 
 on the parameters in our framework. 
Since the hidden scalar $\chi$ has the properties 
 similar to the Higgs boson, 
 we can use the current experimental limits 
 of the Higgs boson search to constrain model parameters. 
The most severe constraint is provided by the Higgs boson search 
 in the two-photon decay mode at Tevatron 
 with the integrated luminosity 1 fb$^{-1}$, 
 which is found to be 
 $ \sigma_h {\rm Br}( h \to \gamma \gamma) \lesssim 0.1$ pb 
 for a light Higgs boson with a mass around 120 GeV 
 \cite{Mrenna:2000qh}. 
Here, $\sigma_h$ is the Higgs boson production cross section 
 at Tevatron, which is dominated by the gluon fusion process.  
The Standard Model predicts 
 $ \sigma_h {\rm Br}( h \to \gamma \gamma) \sim 10^{-3}$ pb, 
 far below the bound. 
However, when this bound is applied to 
 the $\chi$ production with $\Lambda \simeq 1$ TeV, 
 we obtain a severe constraint on the model parameters. 
Comparing the couplings between gluons and $\chi$ 
 to the one between gluons and the Higgs boson in the SM, 
 we find the ratio of the production cross sections 
 at Tevatron as $\sigma_\chi/\sigma_h \sim 100 c_{gg}^2$. 
For $m_\chi=120$ GeV, for example, 
 the main decay mode of the $\chi$ will be 
 into two gluons and photons, 
 and the branching ratio into two photons is estimated as 
 ${\rm Br}(\chi \to \gamma \gamma) \simeq (c_{\gamma \gamma}/c_{gg})^2/9$. 
When we assume the universal coupling $c_1=c_2=c_3$ 
 (equivalent to  $c_{gg}=c_{\gamma \gamma}=c_{WW}=c_{ZZ}$ 
  and $c_{Z \gamma}=0$), 
  the Tevatron bound leads us to $c_1=c_2=c_3 \lesssim 0.1$. 
However, in this case, the $\chi$ production cross section 
 becomes two orders of magnitude smaller than 
 the Higgs boson production cross section at the ILC.

There are many possible choices of 
 the parameter set ($c_1$, $c_2$ and $c_3$) 
 so as to satisfy the Tevatron bound, 
 while keeping the $\chi$ production cross section 
 to be comparable to the Higgs boson one. 
To simplify our discussion, in this paper, 
 we choose a special parameter set: 
 $c_1=c_2=1$ and $c_3=0$, 
 namely the gluophobic but universal for $c_1$ and $c_2$. 
Therefore, the $\chi$ production channel through 
 the gluon fusion is closed. 
For $m_\chi < 2 m_W$, 
 the hidden scalar has a 100\% branching ratio into two photons.

\section{Monte Carlo Simulation}
As estimated in the previous section, 
 if the cutoff is around 1 TeV, 
 the production cross section of the hidden scalar 
 can be comparable to the Higgs boson production 
 cross section at the ILC. 
There are two main production processes 
 associated with a $Z$-boson or a photon. 
In the following, we investigate each process. 
In our analysis, we take the same mass for 
 the hidden scalar and the Higgs boson: 
 $m_\chi=m_h=120$ GeV, as a reference. 

\subsection{Observables to be measured} 
The associated hidden scalar production 
 with a $Z$-boson is very similar to 
 the Higgs production process 
 and their production cross sections are comparable  
 for $\Lambda \simeq 1$ TeV. 
One crucial difference is that the hidden scalar couples 
 to $Z$-bosons through Eq.~(\ref{intgauge2}) 
 so that the $Z$-boson in the final state is mostly transversely polarized. 
On the other hand, in the Higgs boson production 
 the interaction between the Higgs boson and 
 the longitudinal mode of the $Z$-boson dominates. 
In order to distinguish the hidden scalar from the Higgs boson, 
 we will measure 

\noindent
 (1) the angular distribution of the $Z$-boson in the final state,  

\noindent
 (2) the polarization of the $Z$-boson in the final sate. 

\noindent
As shown in the previous section, 
 the branching ratio of the hidden scalar decay 
 is quite different from the Higgs boson one. 
In our reference parameter set,  
 the hidden scalar decays 100\% into two photons.  
The Higgs boson with $m_h=120$ GeV dominantly decays 
 into a bottom and anti-bottom quark pair. 
In order to distinguish the hidden scalar from the Higgs boson, 
 we will measure 

\noindent
 (3) the branching ratios into two photons 
     and into the bottom and anti-bottom quark pair 
     through b-tagging.   

%
%
%

The associated hidden scalar production 
 with a photon is unique and such a process 
 for the Higgs boson  is negligible. 
We will investigate similar things as in the $Z$-boson case.

\subsection{Analysis Framework} 
For Monte Carlo simulation studies of 
 the hidden scalar productions and decays,
 we have developed event generators of the processes: 
 $e^+e^- \to \gamma \chi$ and $e^+e^- \to Z \chi$ 
followed by the $\chi \to \gamma \gamma$ decay, 
 which are now included in
\texttt{physsim-2007a}~\cite{PHYSSIM}.
In the helicity amplitude calculations, we retain the $Z$-boson
 wave function if any and replace it with the wave function 
 composed with the daughter fermion-antifermion pair 
 according to the HELAS algorithm~\cite{Murayama:1992gi}.
This allows us to properly take into account 
 the gauge boson polarization effects. 
The phase space integration and generation of parton 4-momenta 
 are performed with \texttt{BASES/SPRING}~\cite{Kawabata:1985yt}.
Parton showering and hadronization are carried out
 using \texttt{PYTHIA6.3}~\cite{Sjostrand:1993yb} 
 with final-state tau leptons treated 
 by \texttt{TAUOLA}~\cite{Jadach:1993hs} 
 in order to handle their polarizations properly. 
The background $e^+e^- \to Z h$ events are generated 
 using the $e^+ e^- \to Z \chi$ generator
with the $e^+ e^- \to Z \chi$ helicity amplitudes replaced by
corresponding $e^+ e^- \to Z h$ amplitudes
and the Higgs decay handled by \texttt{PYTHIA6.3}.

In the Monte Carlo simulations, we set 
 the nominal center-of-mass energy at $500$ GeV
 and assume no beam polarization. 
Effects of natural beam-energy spread and beamstrahlung 
 are taken into account 
 according to the beam parameters given in \cite{Ref:GLDDOD}.
We have assumed no crossing angle between 
 the electron and the positron beams 
 and ignored the transverse component of the initial state radiation.
Consequently, 
 the $Z \chi$ or $\gamma \chi$ system in our Monte-Carlo sample
 has no transverse momentum.

The generated Monte-Carlo events were passed to a detector 
 simulator (\texttt{JSF Quick Simulator}~\cite{JSF}) 
 which incorporates the ACFA-LC study parameters
 (see Table.~\ref{det-param}). 
The quick simulator created vertex-detector hits, 
 smeared charged-track parameters in the central tracker 
 with parameter correlation properly taken into account,
 and simulated calorimeter signals as from individual segments,
 thereby allowing realistic simulation of cluster overlapping.
It should also be noted that track-cluster matching was 
 performed to achieve the best energy-flow measurements. 

\begin{table}[h]
 \begin{center}
   \begin{tabular}{|c||r|r|} \hline
     \textbf{Detector} & \textbf{Performance} & \textbf{Coverage} \\
     \hline
     \hline
     Vertex detector                                                      &
       $ \sigma_{\mathrm{b}}
         = 7.0 \oplus (20.0/p) \, / \, \sin^{3/2}\theta \; \mu\mathrm{m}$ &
       $ | \cos\theta | \leq 0.90 $ \\
     \hline
     Central drift chamber                                                &
       $ \sigma_{p_{T}}/p_{T}
         = 1.1 \times 10^{-4} p_{T} \> \oplus \> 0.1 \; \% $              &
       $ | \cos\theta | \leq 0.95 $ \\
     \hline
     EM calorimeter                                                       &
       $ \sigma_{E}/E
         = 15 \; \% \> / \sqrt{E} \> \oplus 1 \; \% $                     &
       $ | \cos\theta | \leq 0.90 $ \\
     \hline
     Hadron calorimeter                                                   &
       $ \sigma_{E}/E
         = 40 \; \% \> / \sqrt{E} \> \oplus 2 \; \% $                     &
       $ | \cos\theta | \leq 0.90 $ \\
     \hline
   \end{tabular}
 \end{center}
 \caption[]{
       ACFA study parameters for an LC detector, where
       $p$, $p_{T}$, and $E$ are measured in units of GeV.
       }
 \label{det-param}
\end{table}%

\subsection{Event Selection and Results}
\subsubsection{$e^+e^- \to Z\chi; \chi \to \gamma\gamma$ process} 

Data equivalent to 50 fb$^{-1}$ have been generated for 
 both $e^+e^- \to Z\chi$ followed by $\chi \to \gamma\gamma$ and 
 $e^+e^- \to Zh$ followed by $h \to \gamma\gamma$. 
A typical event is displayed in Figure~\ref{EvtDis_2a2j}. 
For the $Z\chi \to q\bar{q}\gamma\gamma$ process, 
 there are two jets and two photons in the final state. 
In the event selection, it is firstly required 
 that the number of reconstructed particles ($N_{particles}$) 
 is greater than 4. 
In the next, the number of photons reconstructed in the 
 calorimeters ($N_{gammas}$) is greater than 2, 
 and the two photons whose invariant mass 
 is the closest to $m_{\chi}$ are selected. 
Finally, the number of jets ($N_{jets}$) is required to be equal to 2.
These selection criteria are summarized in Table~\ref{Table_2a2j}
 together with efficiency of each cut. 
The distribution of the invariant mass of the two photons 
 which are considered to come from a $\chi$ decay 
 is shown in Figure~\ref{Fig_2a2j} 
 after imposing all the above selection criteria. 
In the figure, as a reference, we also plot the grey histogram 
 corresponding to the $e^+e^- \to Zh$ process 
 with the SM Higgs branching ratio 
 where the number of remaining events is much less than 
 that of the $e^+e^- \to Z \chi$ process.  
However, note that the large number of evens from 
 the $\chi$ decay does not directly means that the $\chi$ 
 has nothing to do with the electroweak symmetry breaking. 
As we discussed in the previous section, some class of 
 new physics models can enhance the branching ratio 
 of $h \to \gamma \gamma$ and so the number of events. 
If this is the case, the grey histogram becomes higher. 
Therefore, for the discrimination, it is crucial 
 to measure the angular distribution and polarization 
 of the final states. 
Figures~\ref{Fig_zdecay} and \ref{Fig_hdecay} show 
 the $\chi$ and Higgs production angles (left) and 
 the angular distribution of the reconstructed jets 
 from associated $Z$-boson decays (right) 
 for the both processes, respectively. 
 As can be seen from these plots, $\chi$ couples 
 with the transverse modes of the $Z$-bosons, 
 while the Higgs boson couples with the longitudinal modes.
The $e^+e^- \to Zh$ followed by $h \to A^0 A^0$ process 
 is also analyzed with the same cut conditions 
 and its cut statistics is summarized in Table~\ref{Table_2a2j}. 
Here, we have assumed ${\rm Br}(h \to A^0 A^0)=0.1$ 
 and ${\rm Br}(A^0 \to \gamma \gamma)=1$. 
The distribution of the invariant mass of the two photons
will be similar to Figure~\ref{Fig_2a2j} in this model,
but again we can discriminate the $\chi$ from the Higgs 
 by looking at the angular distributions.
Figure~\ref{Fig_aadecay} shows the Higgs production angle 
 and the angular distribution of the reconstructed jets 
 from associated $Z$-boson decays (right) 
 for the $h \to A^0 A^0$ process. 


\begin{table}[h]
 \begin{center}
   \begin{tabular}{|c||c|c|c|} \hline
     Cut & $Z\chi; \chi \to \gamma \gamma$ 
         & $Zh; h \to \gamma \gamma$ 
         & $Zh; h \to A^0 A^0$\\ \hline               
     \hline No Cut                                                  
     & 2187 (1.0000) & 142 (1.000) & 7087 (1.0000) \\
     \hline $N_{particles} \ge 4$      
     & 1738 (0.7947) & 106 (0.747) & 5692 (0.8032) \\
     \hline $N_{gammas} \ge 2$ 
     & 1521 (0.8751) & 96 (0.906) &  4865 (0.8547) \\
     \hline Cut on $M_{\gamma\gamma}$      
     & 1499 (0.9855) & 95 (0.990) &  4828 (0.9924) \\
     \hline $N_{jets} = 2$ for Ycut = 0.004 
     & 1498 (0.9993) & 95 (1.000) &  4825 (0.9994) \\
     \hline \hline
     Total Efficiency & $0.6850 \pm 0.0099$ 
     & $0.669 \pm 0.040$ & $0.6808 \pm 0.0055$ \\
     \hline
   \end{tabular}
 \end{center}
 \caption[]{
Cut statistics and breakdown of selection efficiency. 
The numbers inside and outside of parenthesis are the efficiency 
 and the remaining number of events after each cut, respectively. 
Here, we have assumed ${\rm Br}(h \to A^0 A^0)=0.1$ 
 and ${\rm Br}(A^0 \to \gamma \gamma)=1$. 
}
 \label{Table_2a2j}
\end{table}%

\subsubsection{$e^+e^- \to \gamma\chi; \chi \to \gamma\gamma$ process}

Data equivalent to 5.7 fb$^{-1}$ have been generated for both signal 
 ($e^+e^- \to \gamma\chi$ followed by $\chi \to \gamma\gamma$) 
 and background ($e^+e^- \to \gamma\gamma$ with an ISR photon) processes.  
A typical signal event is displayed in Figure~\ref{EvtDis_3a}. 
For the $\gamma\chi \to \gamma\gamma\gamma$ process, 
 there are three photons in the final state. 
The number of photons reconstructed in the calorimeters ($N_{gammas}$) 
 is required to be equal to 3. 
It is also required that 
 the energy and the cosine of the polar angle of each photon 
 are greater than 1 GeV and less than 0.999, respectively.
Among the photons, two photons whose invariant mass 
 is within $m_{\chi} \pm 25$ GeV are considered 
 to be from a $\chi$ decay. 
Finally, the cosines of the production angles 
 of both $\chi$ and the remaining photon are required 
 to be less than 0.99.
These selection criteria are summarized in Table~\ref{Table_3a}
 together with their efficiencies. 
The distribution of the invariant mass of two photons 
 which are considered to come from a $\chi$ decay (left) 
 and the angular distribution of the $\chi$ (right) 
 are shown in Figure~\ref{Fig_3a} after imposing 
 all the above selection criteria.
A peak at $m_{\chi}$ can be clearly seen 
 over the grey background histogram 
 with the angular distribution consistent with 
 $1+\cos^2\theta$.

\begin{table}[h]
 \begin{center}
   \begin{tabular}{|c||c|c|} \hline
 Cut & $\gamma \chi; \chi \to \gamma \gamma$ & 
 $\gamma \gamma$ with an ISR \\ \hline                                
     \hline No Cut                                                  
     & 600 (1.0000) & 100000 (1.0000) \\
     \hline $N_{gammas} = 3$                        
     & 575 (0.9583) & 3746 (0.0375) \\
     \hline $E_{gamma} > 1$ GeV                                           
     & 575 (1.0000) & 3730 (0.9959) \\ 
     \hline $|\cos(\theta_j)| \le 0.999$                            
     & 575 (1.0000) & 3728 (0.9992)\\ 
     \hline $|M_{\gamma\gamma} - m_{\chi}| \le 25$ GeV                             
     & 573 (0.9965) & 1332 (0.3573) \\
     \hline $|\cos(\theta_{\chi})|$ and $|\cos(\theta_a)| \le 0.99$ 
     & 572 (0.9983) & 1269 (0.9529) \\
     \hline \hline
     Total Efficiency & $0.9533 \pm 0.0086$ & $0.0127 \pm 0.0001$ \\
     \hline
   \end{tabular}
 \end{center}
 \caption[]{
Similar to Table~2 
 for $e^+e^- \to \gamma\chi$ and 
 $e^+e^- \to \gamma\gamma$ with an ISR photon. 
}
 \label{Table_3a}
\end{table}%

\section{Summary and discussions} 
If a hidden scalar field appears in a certain class of  
 new physics models around the TeV scale, 
 there are interesting implications 
 for collider phenomenology. 
In particular, since the scalar behaves like 
 the Higgs boson in its production process, 
 it is an interesting issue how to distinguish 
 the scalar from the Higgs boson 
 in future collider experiments. 
We investigated the hidden scalar production at the ILC 
 and addressed this issue based on realistic Monte Carlo simulations. 

With the $\chi$ production cross section comparable 
 to the Higgs boson one, 
 the invariant mass distribution reconstructed from 
 two-photon final states due to the decay mode 
 $\chi \to \gamma \gamma$ shows a clear peak at $m_\chi$. 
In the $\chi$ production associated with a $Z$-boson, 
 the $\chi$ production angle and the angular distribution 
 of the reconstructed jets from the associated $Z$-boson decay 
 reveal that the hidden scalar couples to 
 transversally polarized $Z$-bosons. 
On the other hand, the Higgs boson production 
 associated with a $Z$-boson shows clearly different results 
 in angular distributions and distinguishable 
 from the hidden scalar production.

Some comments are in order here. 
In this paper, we have assumed that the hidden scalar couples 
 with only the SM gauge bosons. 
In general, one can introduce couplings 
 between the hidden scalar and the SM fermions and 
 the Higgs doublets. 
Once these couplings are introduced, 
 the phenomenology of $\chi$ productions can be drastically 
 altered from those in this paper. 
In particular, the coupling of the hidden scalar to 
 the Higgs doublets induces the mixing 
 between $\chi$ and the Higgs boson. 
This mixing spoils the crucial difference 
 between the hidden scalar and the Higgs boson 
 that the former has nothing to do with the electroweak symmetry 
 breaking while the latter is crucial for it. 
It is not natural but in practice, we can assume 
 the above unwanted couplings to be small. 

In fact, it is easy to introduce a setup 
 where the couplings are naturally suppressed. 
As a simple example, let us consider a model 
 in the context of the brane world scenario, 
 where there are two different branes with 
 the three spatial dimensions separated 
 in extra-dimensional directions. 
Suppose that the SM gauge bosons live in the bulk 
 and the hidden scalar resides on one brane 
 while the SM fermions and the Higgs doublets 
 on the other brane. 
In this setup, the couplings between the hidden scalar 
 and the SM fermions and Higgs doublets 
 are geometrically suppressed, while the hidden scalar couples 
 with the bulk SM gauge bosons. 
However, considering loop corrections through gauge bosons, 
 the mixing term between Higgs doublet ($H$) and $\chi$ 
 can be induced which is roughly estimated as 
\bea 
  {\cal L}_{int} \sim 
 \frac{\alpha_2}{4 \pi} \Lambda \chi (H^\dagger H), 
\eea 
where $\alpha_2$ is the SM SU(2) gauge coupling 
 and we have cut off the quadratic divergence 
 of this loop calculation by $\Lambda$. 
After the Higgs doublet develops the VEV, 
 a mass mixing between $\chi$ and the Higgs boson arises. 
Mixing angle is roughly estimated as 
 $\frac{\alpha_2}{4 \pi} \Lambda v/m_\chi^2 \sim 0.05$ 
 for the Higgs VEV $v=246$ GeV, $\Lambda \simeq 1$ TeV 
 and $m_\chi=120$ GeV. 
Thus, only the 5\% of the Higgs boson couples to the transverse mode 
 of the $Z$-boson (only the 5\% of $\chi$ couples to the longitudinal mode 
 of the $Z$-boson), and there is no great importance for our purpose 
 of this paper, though it would be rather interesting 
 how accurately the ILC can measure this mixing. 
While the modification of the $\chi$ decay due to 
 the 5\% mixing (at least, for the parameters we chose) 
 is negligible, this mixing can greatly enhance the Higgs branching 
 into two photons as in the models with $A_0$.

In our analysis, we have taken a special parameter set 
 for simplicity, namely, the gluophobic ($c_3=0$) 
 and universal ($c_2=c_1$) couplings considering the Tevatron bound. 
In general, it is not necessary to take $c_3=0$ in order 
 to avoid the Tevatron bound. 
For example, a parameter set, 
 $c_{gg}=c_{\gamma \gamma}=0.1$ and $c_{ZZ}=1$, 
 can be consistent with the Tevatron bound 
 while keeping the $\chi$ production cross section 
 comparable to that of Higgs boson. 
In this case, $c_{Z \gamma} \simeq 0.7$ and 
 the hidden scalar decays into $Z \gamma$ 
 with a sizable branching ratio. 
It is interesting to study the $\chi$ production 
 through its decay into $Z \gamma$. 
Also, in this parameter set, the decay mode 
 into two gluons is sizable. 
It is hence an interesting issue how to distinguish $\chi$ 
 from the Higgs boson through their hadronic decay modes. 
As can be seen from Eq.~(\ref{dcrossZ}), 
 the coupling between the hidden scalar and 
 the longitudinal mode of the $Z$-boson 
 is proportional to $m_Z/E_Z$ and is sizable at low energy. 
To measure this coupling through the energy-dependence 
 of the production angle distribution 
 may provide an additional handle to pin down 
 the $\chi$ production. 
For this purpose, the ILC with low energy could be useful.  

In this paper, we have concentrated on the hidden scalar 
 production associated with a $Z$-boson or a photon. 
It is also interesting to investigate 
 the weak boson fusion process. 
For example, in the $Z$-boson fusion process, 
 measuring the correlations between the cross section 
 and the azimuthal angle between the final state electron and positron 
 can be used to distinguish the couplings between 
 a scalar and the $Z$-boson with different polarizations.

\noindent {\large \bf Acknowledgments}

This work is supported in part by 
 the Creative Scientific Research Grant (No. 18GS0202)  
 of the Japan Society for Promotion of Science 
 (K.F., H.H., H.I. and T.Y.) and 
 the Grant-in-Aid for Scientific Research 
 from the Ministry of Education, Science and Culture of Japan,
 No. 18740170 (N.O.). 

\def\plb#1#2#3{{\it Phys.~Lett.~}{\bf B#1}, #2 (#3)}
\def\prd#1#2#3{{\it Phys.~Rev.~}{\bf D#1}, #2 (#3)}
\def\prl#1#2#3{{\it Phys.~Rev.~Lett.~}{\bf #1}, #2 (#3)}
\def\zpc#1#2#3{{\it Z.~Phys.~}{\bf C#1}, #2 (#3)}



\newpage
\begin{figure}
\begin{center}
\leavevmode
  \scalebox{1.2}{\includegraphics*{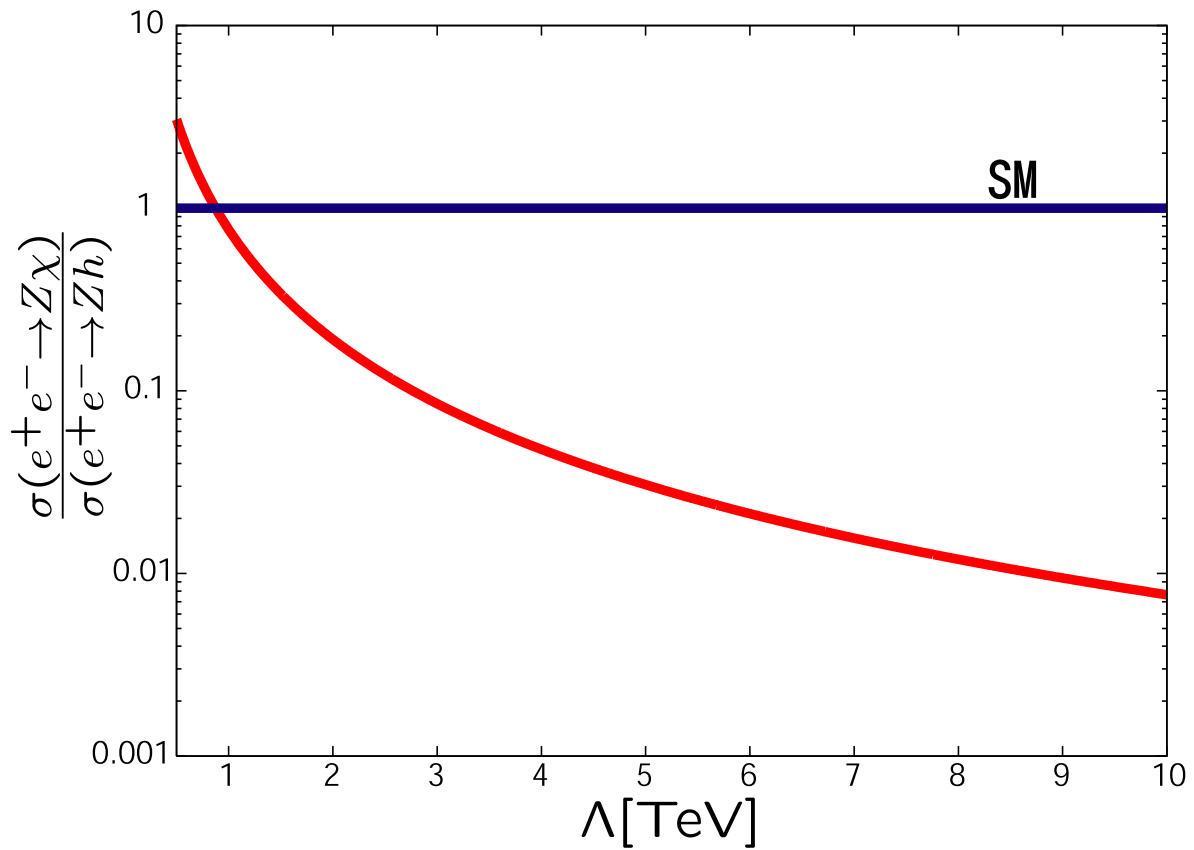}}
\caption{
The ratio of total cross sections 
 between the associated $\chi$ and Higgs productions 
 as a function of $\Lambda$, 
 at the ILC with the collider energy $\sqrt{s} = 500$ GeV. 
Here, we have fixed the parameters such as 
 $m_\chi=m_h=120$ GeV and $c_1=c_2 = 1 $. 
The ratio becomes one for $\Lambda \simeq 872$ GeV. 
}
\end{center}
\end{figure}
\begin{figure}
\begin{center}
\leavevmode
  \scalebox{1.2}{\includegraphics*[width=13cm]{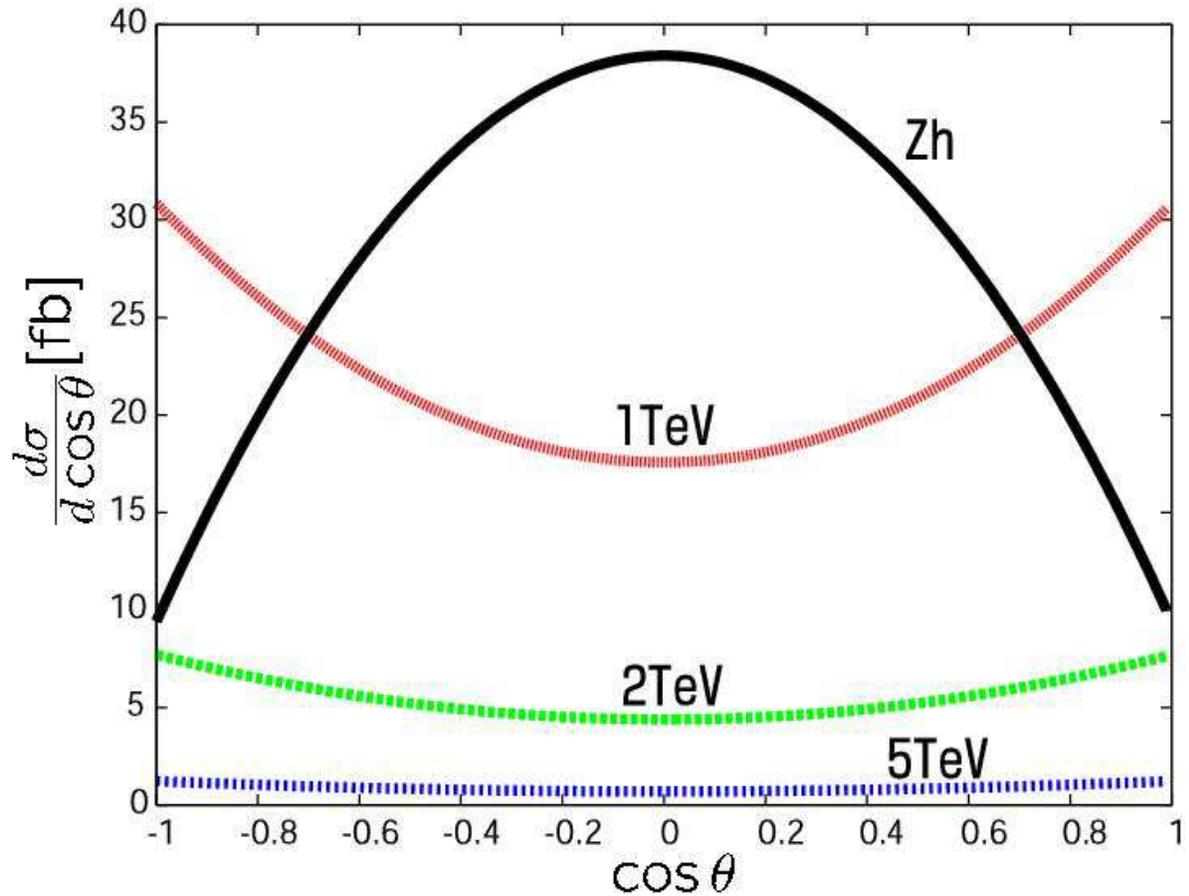}}
\caption{
The angular dependence of the cross sections 
 for $m_\chi=m_h=120$ GeV and $c_1=c_2 =1$, 
 at the ILC with the collider energy $\sqrt{s} = 500$ GeV 
 and $\Lambda =1$, $2$ and $5$ TeV. 
}
\vspace{-0.5cm}
\end{center}
\end{figure}

\begin{figure}
\begin{center}
\leavevmode
  \scalebox{1.2}{\includegraphics*[width=13cm]{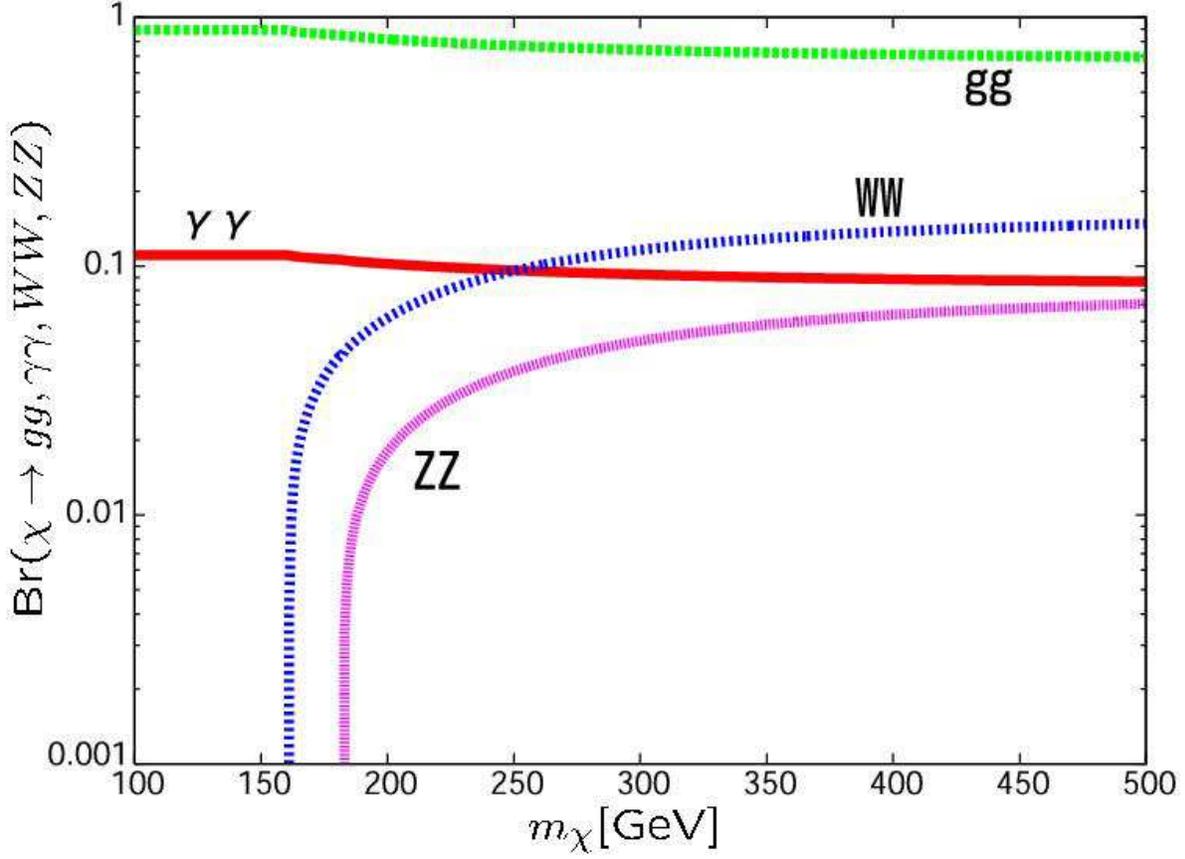}}
\caption{
The branching ratio of the hidden scalar ($\chi$) 
 as a function of its mass $m_\chi$ for 
 $c_1=c_2=c_3=1$. 
Different lines correspond to the modes, 
 $\chi \to gg$, $WW$, $\gamma \gamma$ and $ZZ$. 
}
\end{center}
\end{figure}

\begin{figure}
\begin{center}
\leavevmode
  \includegraphics*[width=8cm]{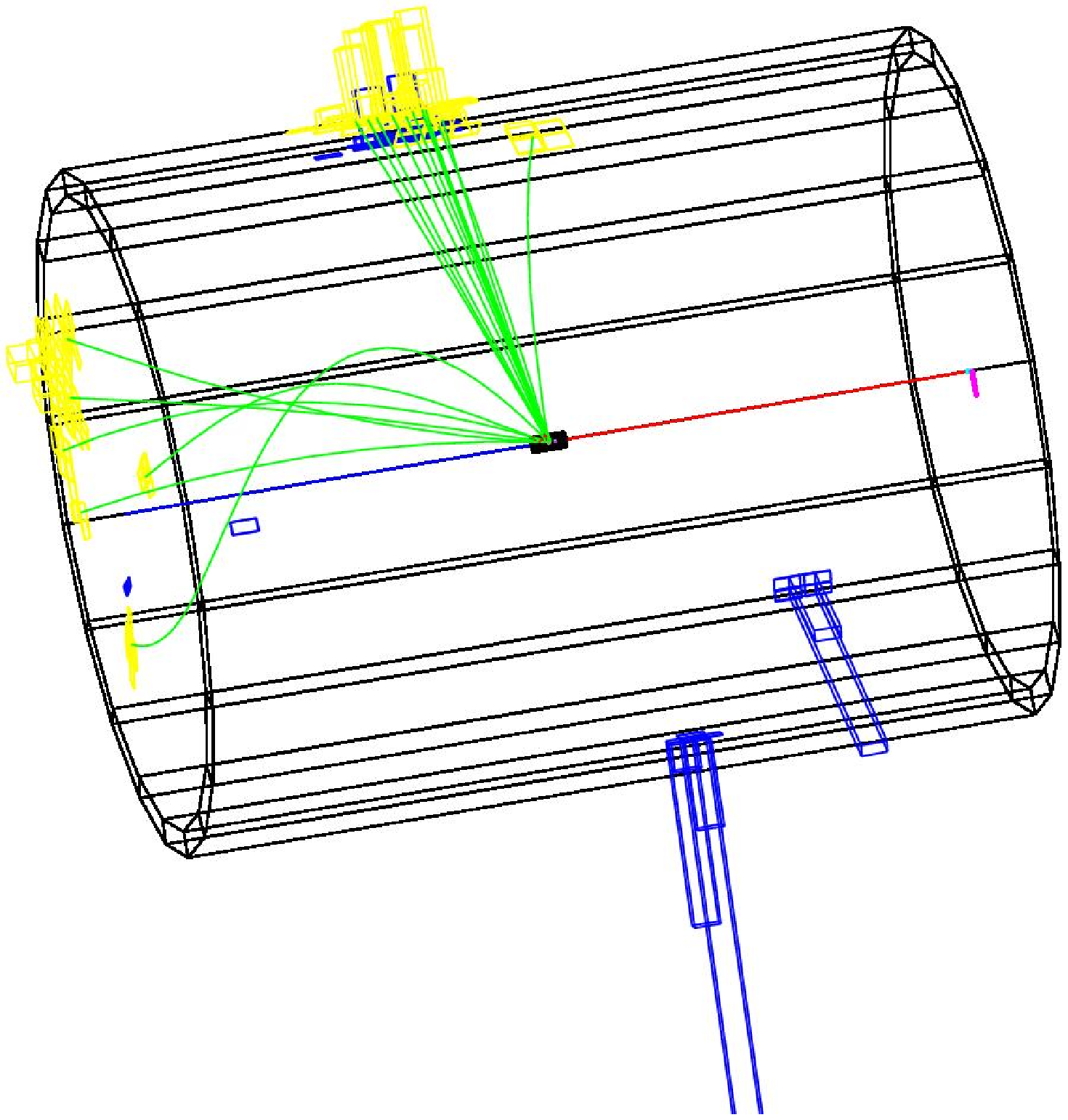}
  \includegraphics*[width=8cm]{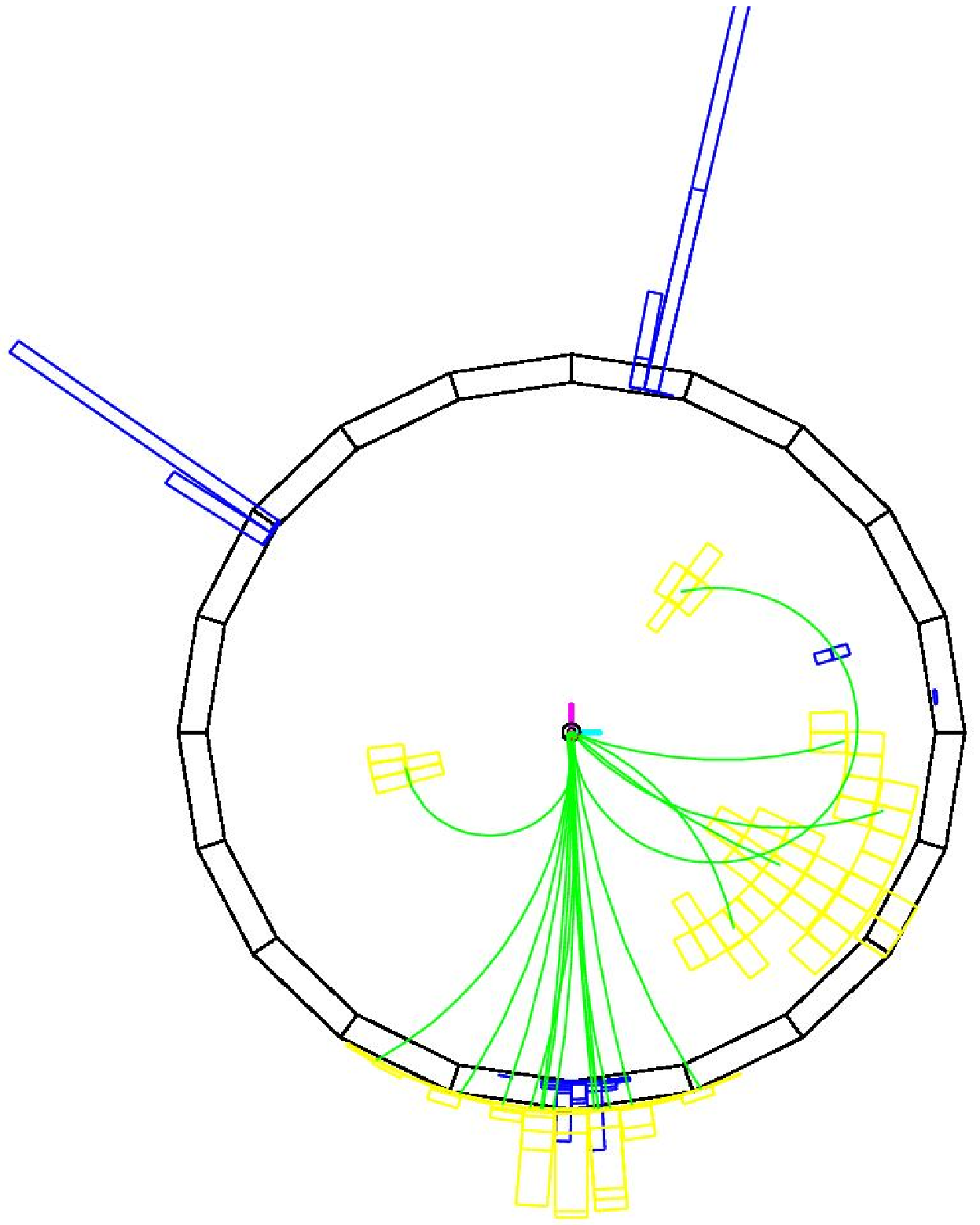}
\caption{
Event displays of $e^+e^- \to Z\chi$ followed by $\chi 
\to \gamma\gamma$. Two jets from the $Z$-boson decay and two photons
from the $\chi$ decay can be clearly seen.}
\label{EvtDis_2a2j}
\end{center}
\end{figure}

\begin{figure}
\begin{center}
\leavevmode
  \scalebox{0.4}{\includegraphics*{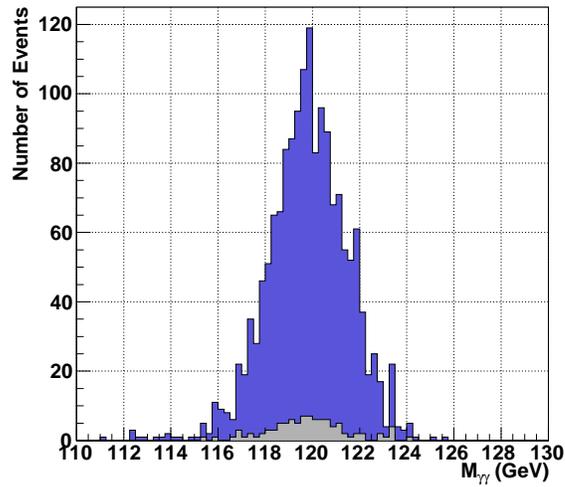}}
\caption{
The distribution of the invariant mass of two photons 
 which are considered to come from a $\chi$ decay.} 
\label{Fig_2a2j}
\end{center}
\end{figure}

\begin{figure}
\begin{center}
\leavevmode
  \scalebox{0.8}{\includegraphics*{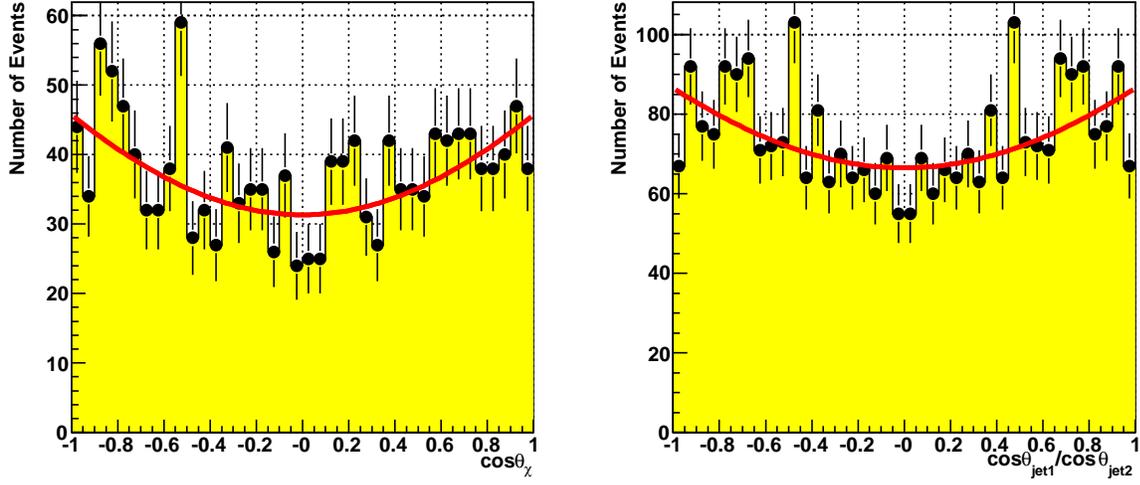}}
  \caption{
The $\chi$ production angle (left) and 
 the angular distribution of the reconstructed jets 
 from associated $Z$-boson decays (right).}
\label{Fig_zdecay}
\end{center}
\end{figure}

\begin{figure}
\begin{center}
\leavevmode
  \scalebox{0.8}{\includegraphics*{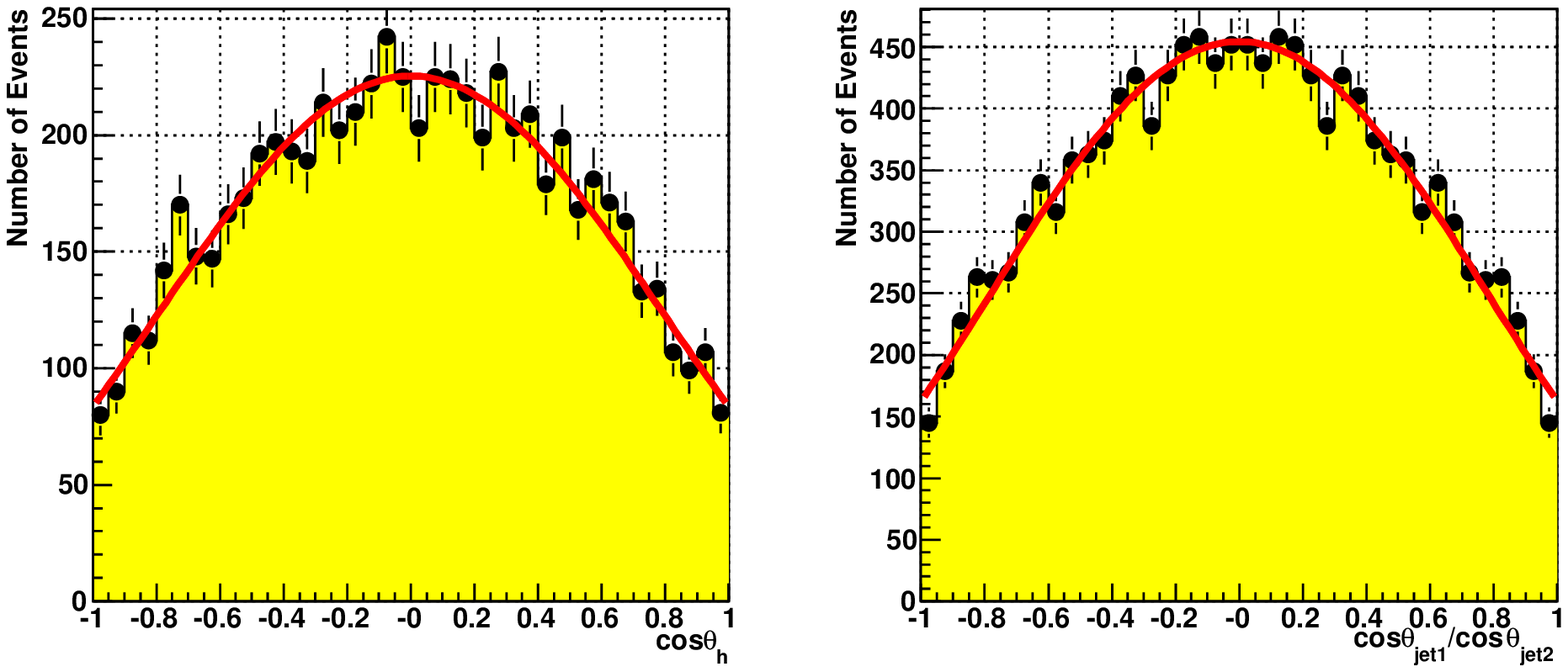}}
  \caption{
The Higgs production angle (left) and 
 the angular distribution of the reconstructed jets 
 from associated $Z$-boson decays (right)
  for $e^+e^- \to Zh$ followed by $H \to \gamma\gamma$. 
}
\label{Fig_hdecay}
\end{center}
\end{figure}

\begin{figure}
\begin{center}
\leavevmode
  \scalebox{0.8}{\includegraphics*{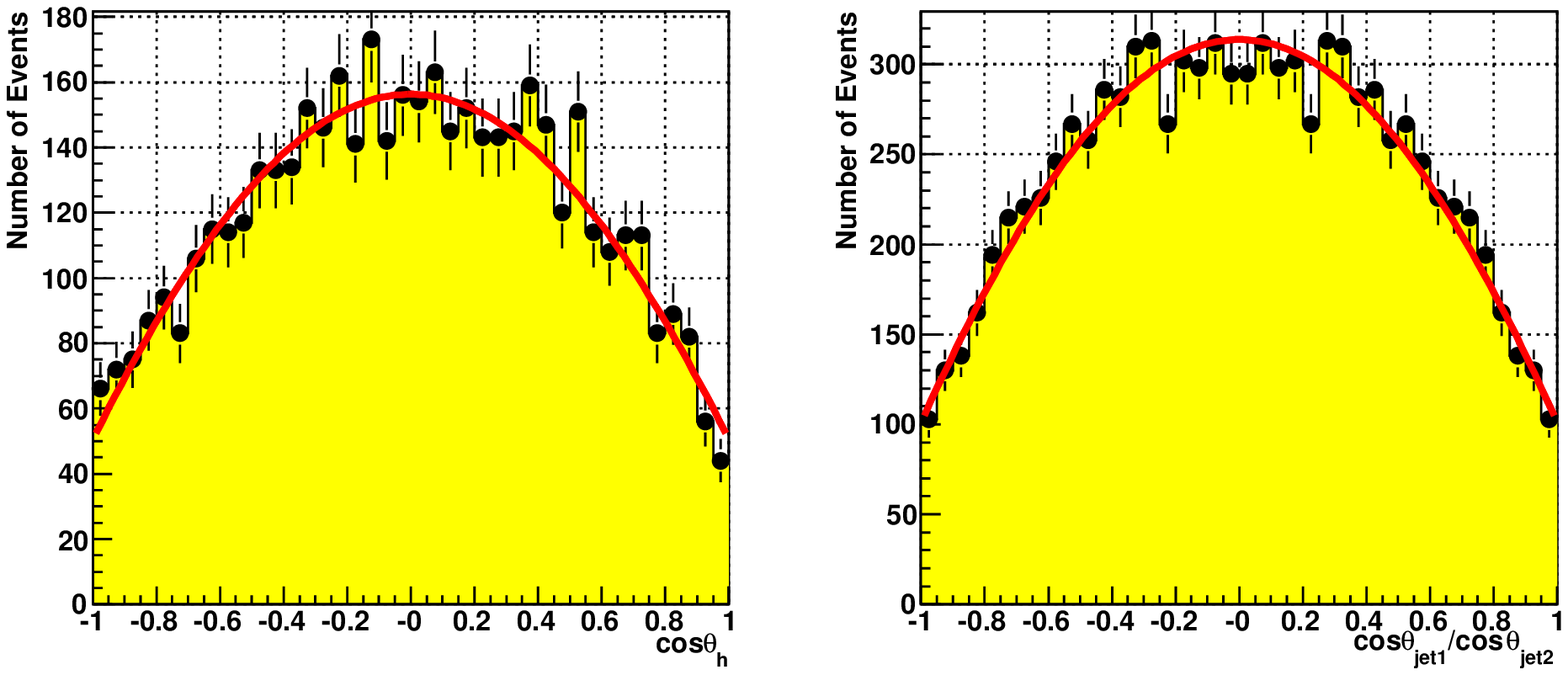}}
  \caption{
The Higgs production angle (left) and 
 the angular distribution of the reconstructed jets
 from associated $Z$-boson decays (right)
  for $e^+e^- \to Zh$ followed by $h \to A^0 A^0 $.
}
\label{Fig_aadecay}
\end{center}
\end{figure}

\begin{figure}
\begin{center}
\leavevmode
  \scalebox{0.4}{\includegraphics*{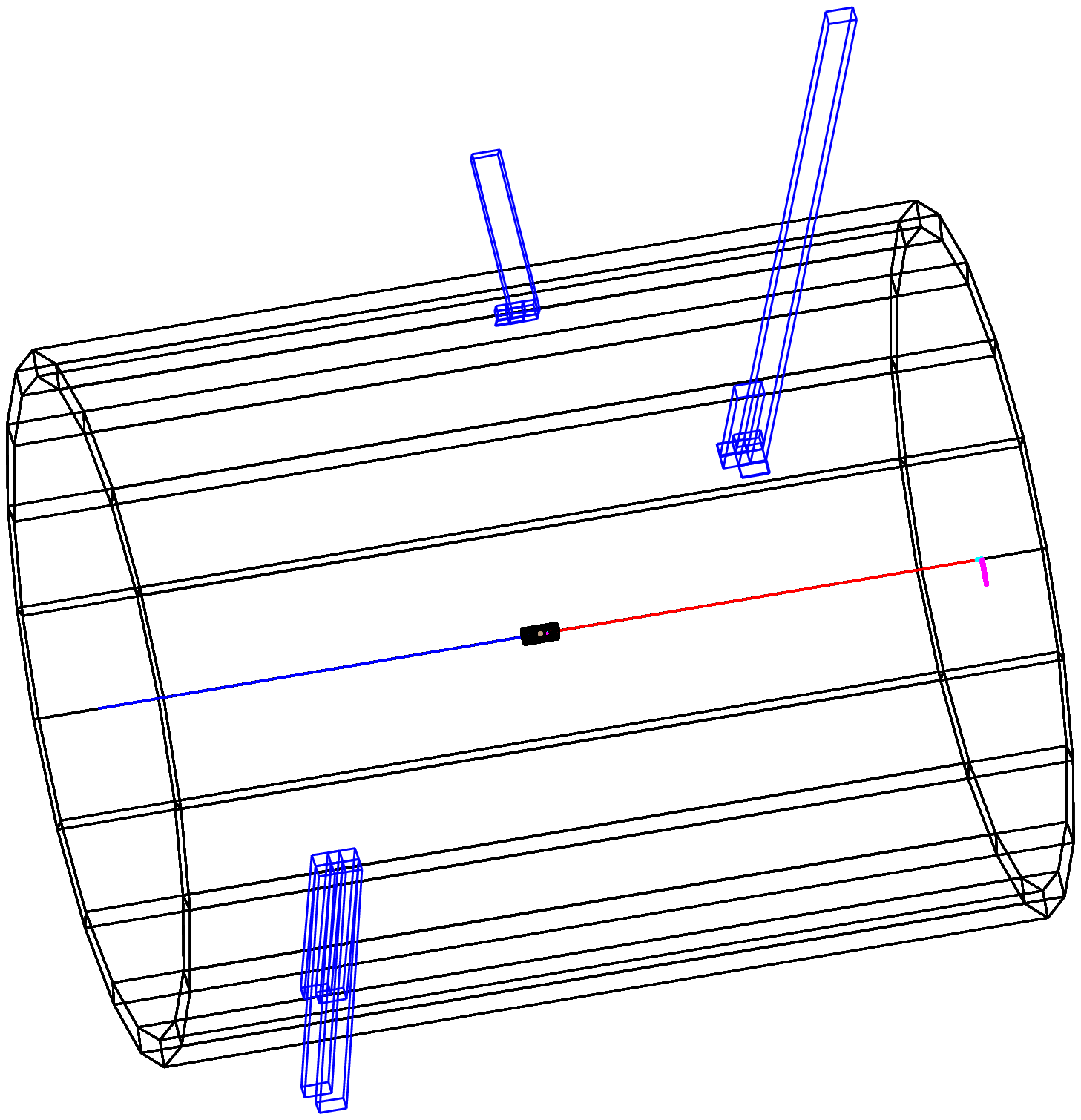}}
  \scalebox{0.4}{\includegraphics*{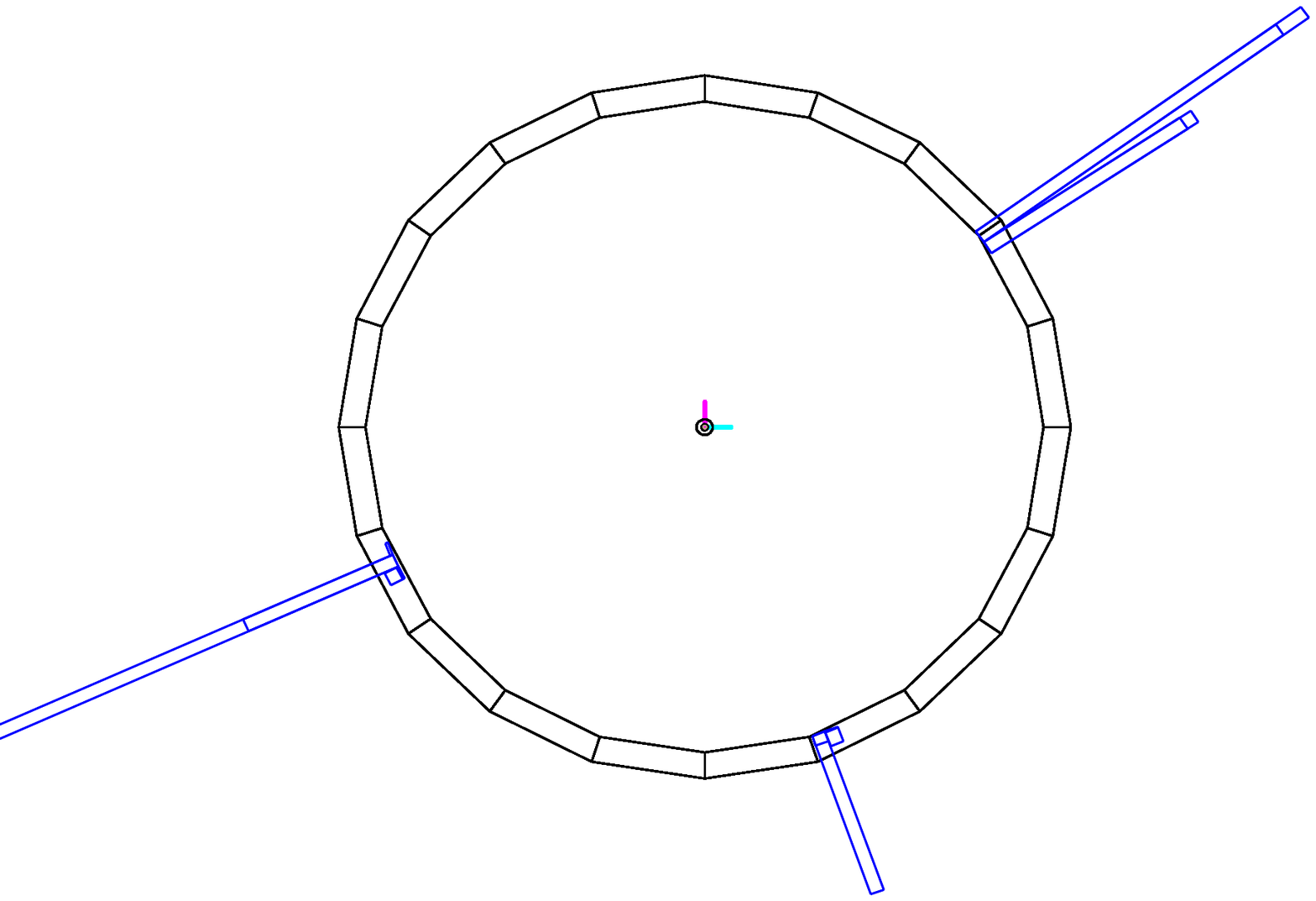}}
\caption{
Event displays of $e^+e^- \to \gamma\chi$ followed by $\chi 
 \to \gamma\gamma$.
}
\label{EvtDis_3a}
\end{center}
\end{figure}

\begin{figure}
\begin{center}
\leavevmode
  \scalebox{0.8}{\includegraphics*{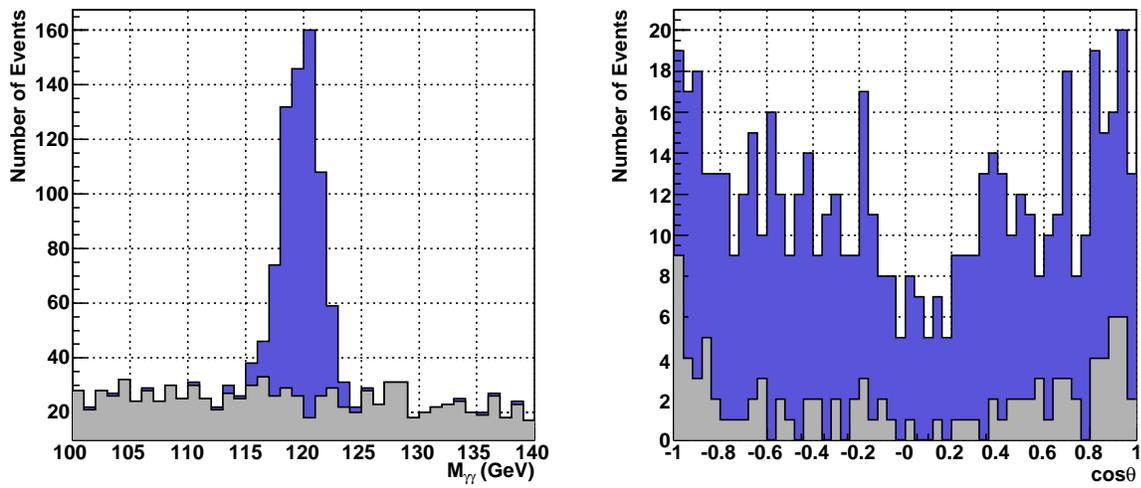}}
\caption{
The distribution of the invariant mass 
 of two photons which are considered to come 
 from a $\chi$ decay (left) and the angular distribution 
 of the $\chi$ (right) for the $e^+e^- \to \gamma \chi$ process
 with background.}
\label{Fig_3a}
\end{center}
\end{figure}

\end{document}